\documentclass[titlepage, 12pt]{article}
\usepackage{amsmath, amsthm, amsfonts, amssymb, amsxtra}
\usepackage{graphicx,psfrag,epsf}
\usepackage{enumerate}
\usepackage{natbib}
\usepackage{url} 
\usepackage{bm, bbm}
\usepackage{booktabs}
\usepackage{hyperref}
\usepackage{xcolor}
\usepackage{subcaption}
\usepackage{physics} 
\usepackage{multirow} 
\usepackage[title]{appendix}
\usepackage{makecell}
\usepackage[justification=centering]{caption} 
\usepackage{blkarray} 
\usepackage{threeparttable} 
\usepackage{arydshln} 
\usepackage{nicematrix}
\DeclareMathOperator{\diag}{diag}
\DeclareMathOperator{\cov}{cov}
\DeclareMathOperator{\variance}{var}
\DeclareMathOperator{\cor}{cor}

\allowdisplaybreaks

\usepackage[margin = 1in]{geometry}
\usepackage{setspace}

\usepackage[]{lineno}

\newcommand*\patchAmsMathEnvironmentForLineno[1]{%
  \expandafter\let\csname old#1\expandafter\endcsname\csname #1\endcsname
  \expandafter\let\csname oldend#1\expandafter\endcsname\csname 
end#1\endcsname
  \renewenvironment{#1}%
  {\linenomath\csname old#1\endcsname}%
  {\csname oldend#1\endcsname\endlinenomath}}%
\newcommand*\patchBothAmsMathEnvironmentsForLineno[1]{%
  \patchAmsMathEnvironmentForLineno{#1}%
  \patchAmsMathEnvironmentForLineno{#1*}}%
\AtBeginDocument{%
  \patchBothAmsMathEnvironmentsForLineno{equation}%
  \patchBothAmsMathEnvironmentsForLineno{align}%
  \patchBothAmsMathEnvironmentsForLineno{flalign}%
  \patchBothAmsMathEnvironmentsForLineno{alignat}%
  \patchBothAmsMathEnvironmentsForLineno{gather}%
  \patchBothAmsMathEnvironmentsForLineno{multline}%
}

\newcommand{\blind}{0}

\newcommand{\E}{\mathbbm{E}}
\newcommand*{\rom}[1]{\expandafter\@slowromancap\romannumeral #1@}
\newcommand{\RNum}[1]{\uppercase\expandafter{\romannumeral #1\relax}}

\graphicspath{ {img/}}
\hypersetup{
    colorlinks = true,
    linkcolor  = blue,
    urlcolor   = blue,
    citecolor  = blue,
    pdftitle   = {Generalized Estimation Equations}
  }

\begin{document}

\title{On GEE for Mean-Variance-Correlation Models: Variance Estimation and Model 
Selection}

\if0\blind
{
  \author{Zhenyu Xu\\
    Department of Statistics, University of Connecticut \\
    Jason P. Fine\\
    Department of Statistics, University of Pittsburgh\\
    Wenling Song\\
    The First Hospital of Jilin University, Changchun, China\\
    Jun Yan\\
    Department of Statistics, University of Connecticut \\
   }
} \fi

\if1\blind
{
  \author{Anonymous Author}
} \fi

\maketitle

\begin{abstract}
  Generalized estimating equations (GEE) are of great importance in 
      analyzing clustered data without full specification of multivariate 
      distributions. A recent approach jointly models the mean, variance, and 
      correlation coefficients of clustered data through three sets of 
      regressions \citep{luo2022conditional}.
      While this method provides a novel framework, we note that 
      it represents a specific case of the more general estimating equations 
      proposed by \citet{yan2004estimating} which further allow 
      the variance 
      to depend on the mean through a variance function. In certain scenarios, 
      the proposed variance estimators for the variance and correlation 
      parameters in \citet{luo2022conditional} may face 
      challenges due to the 
      subtle dependence induced by the nested structure of the estimating 
      equations. We characterize specific model settings where their variance 
      estimation approach may encounter limitations and illustrate how the 
      variance estimators in \citet{yan2004estimating} can 
      correctly account 
      for such dependencies. In addition, we introduce a novel model selection 
      criterion that enables the simultaneous selection of the 
      mean-scale-correlation model. The sandwich variance estimator and the 
      proposed model selection criterion are tested by several simulation 
      studies and real data analysis, which validate its effectiveness in 
      variance estimation and model selection. Our work also extends the R 
      package \texttt{geepack} with the flexibility to apply different working 
      covariance matrices for the variance and correlation structures.

   \vfill
\end{abstract}

\doublespacing

\section{Introduction}
\label{sec:intro}

Clustered data are ubiquitous in various fields,
including economics, public health, and biomedical science. Generalized 
estimating equations (GEE) have become a popular tool for analyzing clustered 
data since its introduction by \citet{liang1986longitudinal}.
The first version 
of GEE (GEE1) avoids the need to specify a joint distribution and yields 
consistent mean parameter estimators, even when the covariance structure is not 
correctly specified. The estimation efficiency is higher when the working 
covariance structure is closer to the true covariance structure. As in a 
generalized linear model (GLM) setting, the response variable's variance is the 
product of a known function of the mean and a scale parameter. This 
variance function serves as an indicator of heteroscedasticity, showing how the 
variance of a random quantity changes in relation to its mean. The 
scale parameter and within-cluster correlation parameters are generally 
estimated using the method of moments. Consequently, the efficiency in 
estimating the mean parameters may be compromised if either the variance or the 
correlation or both are incorrectly specified
\citep{wang2005effects, wang2003working}. In many applications such as family 
studies in genetic epidemiology, the association structure itself may 
be of explicit interest.

Additional estimating equations have been proposed for characterizing the
covariance structure of clustered data beyond marginal means. For instance,
\citet{smyth1989generalized} and \citet{paik1992parametric} introduced
covariate effects onto the variance heterogeneity through a link function and
utilized a second estimating equation for the scale parameters. On the other
hand, \citet{prentice1988correlated} employed a second estimating 
equation (GEE2) to model the within-cluster association structure for binary 
data. \citet{prentice1991estimating} extended the
approach to cover a wider range of general responses beyond binary data.
\citet{carey1993modelling} improved the estimation efficiency using
alternating logistic regressions.
These techniques offer simultaneous inferences for both mean and
association parameters, but the mean parameter estimators may become
inconsistent if the association structure is incorrectly specified.
Alternative methods such as the modified Cholesky decomposition (MCD)
suggested by \citet{pourahmadi1999joint} and the hyperspherical 
coordinates decomposition (HPC) introduced by \citet{zhang2015joint} 
require that the responses in a cluster are naturally ordered, which is often 
not the case in practice. Additionally, the estimated parameters based on MCD 
and HPC cannot be easily interpreted in terms of variances and correlations.

\citet{yan2004estimating} proposed regression models for the mean, 
scale, and
correlation parameters separately and estimated the three sets of parameters
through three estimating equations. The estimating equations have a
hierarchical structure such that the equations for the scale parameters
depend on the mean parameter estimates, and the equation for the correlation
parameters depends on both the mean and scale parameter estimates. Similar to
GEE1, the mean parameter estimators are consistent regardless of
misspecification of the correlation and scale, and the scale parameter
estimators are consistent regardless of misspecification of the correlation. 
The
open-source R package \texttt{geepack} \citep{hojsgaard2006r} implements this
approach. \citet{luo2022conditional} proposed a similar set of three 
estimating equations. The difference is that the variance in \citet{luo2022conditional}
does not include a variance function of the mean as in a GLM setting, so their
approach is a special case of \citet{yan2004estimating} when the 
variance function is constant (as for the Gaussian family). In this case, the 
two methods yield identical point estimates.
However, a key distinction lies in the variance estimation process for these 
point estimates, where they employ different sandwich variance estimators. 
Accurate variance estimates are crucial for statistical inferences and model 
selection.

The contributions of this paper are two-fold. First, we identify situations
where the variance estimator of \citet{luo2022conditional} cannot be 
employed. 
Due to the hierarchical structure of the estimating equations, the bread part 
of the sandwich variance estimator should be a block-triangular matrix while
\citet{luo2022conditional} used a block-diagonal matrix. When the 
off-diagonal 
blocks are nonzero, this can affect the accuracy of their variance estimator. 
This overlook is similar to that in \citet{kastner1999comparison} and
\citet{ziegler2000familial} in a
two-equation setting, and the same amendment of \citet{yan2004estimating}
applies. This potential issue in variance estimation also has implications for 
the reliability of their model selection results. Consequently, our second 
contribution aims to develop a novel 
model selection criterion for estimating equation models. Inspired by the least 
squares approximation of \citet{wang2007unified} in variable 
selection, we 
treat the estimating functions as the gradient of an unknown objective function 
and approximate the objective function using a second-order Taylor expansion at 
the full model estimates. The model selection criterion evaluated for a 
candidate model is the approximated objection function value penalized by a 
measure of the model complexity constructed from the sandwich variance 
estimator of the candidate model, similar to the penalty in 
\citet{pan2001akaike}. The performance of our proposed criterion is 
competitive in our simulation studies.

The rest of the paper is organized as follows. The GEE model as well as its 
updating algorithm and the similarity and difference between the sandwich 
estimator of \citet{yan2004estimating} and \citet{luo2022conditional} are 
presented in Section~\ref{sec:model}. The novel model selection criterion is 
justified in Section~\ref{sec:model_selection}. The performance of the 
different variance estimators and model
selection approaches is evaluated through a series of simulation studies 
outlined in Section~\ref{sec:sim_study}. An application of the proposed methods 
is conducted in Section~\ref{sec:PI_data}. A discussion concludes in 
Section~\ref{sec:discussion}.

\section{Generalized Estimating Equations}
\label{sec:model}

\subsection{Model}

Some notations are first introduced. Consider observations from a sample of $n$ 
independent clusters. Let $Y_i = (Y_{i1}, \ldots, Y_{im_i})^{\top}$ be the
observation vector of cluster~$i$ of size $m_i$, $i = 1, \ldots, n$. Let 
$\mu_{ij}$ and $\sigma_{ij}^2$ be the marginal mean and variance of 
$Y_{ij}$, $j = 1, 2, \ldots, m_i$. Decompose the covariance matrix of $Y_i$ as
\begin{equation}
\label{eq:cov}
  \Sigma_i = \Delta_{1i}^{\frac{1}{2}} R_{1i} \Delta_{1i}^{\frac{1}{2}},
\end{equation}
where $\Delta_{1i} = \diag(\sigma^2_{i1}, \dots, \sigma^2_{i m_i})$ and 
$R_{1i}$ is the correlation matrix of $Y_i$. Under the GLM framework, it is 
often assumed that $\sigma^2_{ij} = \phi_{ij} v_{ij}$, where $\phi_{ij}$ is the 
scale function and $v_{ij}$ is the variance function of the mean response, 
i.e., $v_{ij} = v(\mu_{ij})$.

The GEE model is obtained by introducing covariates into the marginal 
mean, marginal variance, and the pairwise correlation coefficients of 
observation vector $Y_i$. Specifically, let 
$\mu_i = (\mu_{i1}, \ldots, \mu_{im_i})^{\top}$ be the conditional mean vector 
of $Y_i$ given a $m_i \times p$ covariate matrix $X_{1i}$, i.e., 
$\mu_i = E[Y_i|X_{1i}]$. Similarly, let 
$\phi_i = (\phi_{i1}, \ldots, \phi_{im_i})^{\top}$ be the scale parameter 
vector of $Y_i$ given an $m_i\times q$ covariate matrix $X_{2i}$. Let 
$\rho_{ijk}$ be the $(j,k)$th entry of correlation matrix $R_i$ and stack the 
upper-triangular entries of $R_{1i}$ to form an $m_i(m_i-1)/2 \times 1$ vector 
$\rho_i = (\rho_{i12}, \ldots, \rho_{i (m_i-1) m_i})^{\top}$. Let $X_{3i}$ be 
an $m_i(m_i-1)/2 \times q$ covariate matrix for $\rho_i$. A joint regression 
model for the mean, scale, and correlation coefficients is
\begin{align}
  g_1(\mu_i)  &= X_{1i} \beta,    \notag\\
  g_2(\phi_i) &= X_{2i} \lambda,  \label{eq:model}\\
  g_3(\rho_i) &= X_{3i} \gamma,   \notag,
\end{align}
where $g_i$, $i=1, 2, 3$, are known link functions for the mean, the scale and 
the correlation, respectively, and $\beta$, $\lambda$, and $\gamma$ are 
$p\times 1$, $r\times 1$, and $q\times 1$ vectors of regression coefficients 
\citep{yan2004estimating}.

Model~\eqref{eq:model} covers the model of \citet{luo2022conditional} as a
special case where the variance function is a constant $v_{ij} = 1$ as in the
Gaussian family. It can handle over- or under-dispersion and heteroscedasticity
through the scale parameter and the variance function.

\subsection{Estimating Equations}

\citet{yan2004estimating} presented three estimating equations to 
estimate
$\beta$, $\lambda$, and $\gamma$ in~\eqref{eq:model}. Let $s_i$ be the
$m_i\times 1$ vector of $s_{ij} = (y_{ij} - \mu_{ij})^2 / v_{ij}$ and $z_i$
the $m_i (m_i - 1)/2 \times 1$ vector of
$z_{ijk} = (y_{ij} - \mu_{ij}) (y_{ik}-\mu_{ik}) /
\sqrt{\phi_{ij} v_{ij}\phi_{ik} v_{ik}}$.
The estimating equations are
\begin{align}
  \label{eq:beta}
  U_1(\beta, \lambda, \gamma) &=
  \sum_{i=1}^n D_{1i}^{\top} V_{1i}^{-1} (Y_i - \mu_i)=0,\\
  \label{eq:lambda}
  U_2(\beta, \lambda, \gamma) &=
  \sum_{i=1}^n D_{2i}^{\top} V_{2i}^{-1} (s_i - \phi_i)=0,\\
  \label{eq:gamma}
  U_3(\beta, \lambda, \gamma) &=
  \sum_{i=1}^n D_{3i}^{\top} V_{3i}^{-1} (z_i - \rho_i)=0,
\end{align}
where $D_{1i}$ is the $m_i \times p$ matrix of
$\partial \mu_i / \partial \beta^{\top}$,
$D_{2i}$ is the $m_i \times r$ matrix of
$\partial \phi_i / \partial \lambda^{\top}$,
$D_{3i}$ is the $m_i(m_i-1)/2\times q$ matrix of
$\partial \rho_i / \partial \gamma^{\top}$, and $V_{1i}$,
$V_{2i}$, and $V_{3i}$ are, respectively, the conditional working covariance
matrices of $Y_i$, $s_i$
and $z_i$. In the special case of a constant variance function, the estimating
equations are the same as those in \citet{luo2022conditional}.

The working covariance matrices must be determined in order to fit the model 
effectively. The scale parameters $\lambda$ and correlation parameters $\gamma$ 
are often contained in the matrix $V_{1i}$ to approximate the conditional 
version of the covariance matrix~\eqref{eq:cov}. There may be additional 
estimated factors that describe the third and fourth moments in the matrices 
$V_{2i}$ and $V_{3i}$. The formulations used by \citet{paik1992parametric} 
and 
\citet{prentice1988correlated} for $V_{2i}$ for general responses and 
$V_{3i}$ 
for binary responses, respectively, may result in efficiency gains. In 
practice, $V_{2i}$ may be selected to be a diagonal matrix with diagonal 
components $2\phi_{ij}^2$, following the independent Gaussian working matrix
\citep{prentice1991estimating}; and $V_{3i}$ may be an identity matrix, at the 
cost of potential efficiency loss \citep{ziegler1998generalised}. This avoids 
the specification of higher-order moments, estimation of higher-order nuisance 
parameters, and convergence difficulties. An implementation of the estimating
equations~\eqref{eq:beta}--\eqref{eq:gamma} from 
\citet{yan2004estimating} is
available in the R package \texttt{geepack} \citep{hojsgaard2006r}.

Denoting $\theta = (\beta^{\top}, \lambda^{\top}, \gamma^{\top})^{\top}$, we 
estimate $\theta$ by $\hat\theta_n$, the roots of the  estimating 
equations~\eqref{eq:beta}--\eqref{eq:gamma}. This may be obtained by an 
iterative algorithm similar 
to Fisher scoring as proposed by \citet{yan2004estimating}. Each 
component of 
the parameters is updated successively, beginning from a set of sensible 
initial values. Specifically, at step $t+1$ of the iteration,
\begin{align*}
  \hat{\beta}_n^{(t+1)} = \hat{\beta}_n^{(t)} +
  \left( \sum_{i=1}^n D_{1i}^{\top} V_{1i}^{-1}  D_{1i} \right)^{-1}
  \sum_{i=1}^n D_{1i}^{\top} V_{1i}^{-1} (Y_i - \mu_i),\\
  \hat{\lambda}_n^{(t+1)} = \hat{\lambda}_n^{(t)} +
  \left( \sum_{i=1}^n D_{2i}^{\top} V_{2i}^{-1}  D_{2i} \right)^{-1}
  \sum_{i=1}^n D_{2i}^{\top} V_{2i}^{-1} (s_i - \phi_i),\\
  \hat{\gamma}_n^{(t+1)} = \hat{\gamma}_n^{(t)} +
  \left( \sum_{i=1}^n D_{3i}^{\top} V_{3i}^{-1} D_{3i} \right)^{-1}
  \sum_{i=1}^n D_{3i}^{\top} V_{3i}^{-1} (z_i - \rho_i),
\end{align*}
and the second term on the right side of each expression is evaluated at the
most recently updated estimates of the parameters. The initial values can be
obtained from constant scales and working independence. The iteration stops 
when a pre-set convergence tolerance is met. This algorithm reduces to 
\citet{luo2022conditional} under a constant variance function.

In the study by \citet{luo2022conditional}, distinct non-diagonal 
$V_{2i}$ and 
$V_{3i}$ matrices were employed, differing from those mentioned previously. 
They suggested working variances from the multivariate normal distribution with 
$\variance(\epsilon_{ij}^2) = 2(\phi_{ij} v_{ij})^2$ and 
$\variance(z_{ijk}) = 1 + \rho_{ijk}^2$, where 
$\epsilon_{ij} = Y_{ij} - \mu_{ij}$, and constructed $V_{2i}$ and $V_{3i}$ 
similar to $V_{1i}$ in equation~\eqref{eq:cov} by
\begin{align}
  \label{eq:working_lambda}
  {V}_{2i} = \Delta_{2i}^{\frac{1}{2}} R_{2i}(u_1) \Delta_{2i}^{\frac{1}{2}},\\
  \label{eq:working_gamma}
  {V}_{3i} = \Delta_{3i}^{\frac{1}{2}} R_{3i}(u_2) \Delta_{3i}^{\frac{1}{2}},
\end{align}
where $\Delta_{2i} = \diag(2\phi_{i1}^2, \dots, 2\phi_{im_i}^2)$,
$\Delta_{3i} = \diag(1+\rho_{i12}^2, \ldots, 1 +
  \rho_{i1m_i}^2, \dots, 1 + \rho_{i(m_i - 1)m_i}^2)$,
and working correlation matrices $R_{2i}(u_1)$ and $R_{3i}(u_2)$ such as
independence, Order-1 Autoregressive (AR(1)) or Compound Symmetry (CS) 
structures for longitudinal data with fixed working parameters. Interestingly, 
their numerical studies showed that different working correlation matrices in 
$V_{2i}$ and $V_{3i}$ had little impact on the efficiency of the estimation.

\subsection{Sandwich Variance Estimator}

The estimating equations~\eqref{eq:beta}--\eqref{eq:gamma}
have a hierarchical structure that is
important in understanding the properties of the resulting estimator. The
estimating equation~\eqref{eq:lambda} for $\lambda$ depends on $\beta$, and the
estimating equation~\eqref{eq:gamma}
for $\gamma$ depends on both $\beta$ and $\lambda$. Consequently, $\hat\beta_n$
is consistent as long as the mean model is correctly specified; $\hat\lambda_n$
is consistent if both the mean and the scale models are correctly specified;
$\hat\gamma_n$ is only consistent if the mean, scale, and correlation models
are correctly specified. While both \citet{yan2004estimating} and
\citet{luo2022conditional} draw the same conclusion on the 
consistency of
$\hat\theta_n$, their estimators for the variance of $\hat\theta_n$ differ.

The hierarchical structure of the estimating
equations~\eqref{eq:beta}--\eqref{eq:gamma} implies that their slope matrix is
not block diagonal but block lower triangular. The uncertainty in $\hat\beta_n$
affects the variance of both
$\hat\lambda_n$ and $\hat\gamma_n$; the uncertainty in $\hat\lambda_n$ affects
the variance of $\hat\gamma_n$, which is in the same spirit as the slope matrix
in the two-estimating-equation setting of 
\citet{paik1992parametric, prentice1988correlated}.
The variance of $\hat\theta_n$ is of a sandwich
form. The meat matrix is the variance of the estimating functions evaluated at
$\hat\theta_n$. The bread matrix is the inverse of the slope matrix of the
estimating equations. \citet{yan2004estimating} and \citet{luo2022conditional}
have the same meat matrix. The bread matrix of \citet{luo2022conditional}
discarded the block lower triangular feature of the slope matrix and only kept
the diagonal blocks. This is similar to \citet{kastner1999comparison} and
\citet{ziegler2000familial} with two estimating equations, which
was pointed out by \citet{yan2004estimating}. The sandwich 
estimator of \citet{yan2004estimating} is detailed in 
Appendix~\ref{appendix:sand_var}. Its
validity is shown in the simulation study in Section~\ref{sec:sim_study}.

The variance estimator of \citet{luo2022conditional} is 
approximately valid only
in special settings where the off-diagonal blocks of the lower-triangular slope
matrix have expectation zero. In fact, in the simulation studies of 
\citet{luo2022conditional}, their variance estimates agree with the 
empirical variance closely. One can show that in these simulation settings, the 
off-diagonal blocks of the lower-triangular slope matrix have expectations 
close to zero. This generally occurs when the variance function is constant, 
leading to $\dd v_{ij} / \dd \mu_{ij}$ being zero. Consequently, this results 
in zero expectations for two out of the three off-diagonal blocks. Regarding 
the third block, when the average of $\rho_{ijk}$ across all values of $i$, 
$j$, and $k$ approaches zero, its expectation also tends towards zero. If all 
three off-diagonal blocks have zero expectations, then the variance estimator 
proposed by \citet{luo2022conditional} becomes approximately valid. 
Additional 
details on this topic are provided in Appendix~\ref{appendix:coincidence}. 
In our simulation studies in Section~\ref{sec:sim_study}, we employ
scenarios where these off-diagonal blocks have nonzero expectations, in which
case, the variance estimate of \citet{luo2022conditional} does not 
agree with
the empirical variance and the resulting confidence interval does not have the
desired coverage rate. The variance estimator of 
\citet{yan2004estimating}
remains valid in all scenarios, and is used in the model selection criterion 
proposed in the next section.

\section{Model Selection}
\label{sec:model_selection}

In this section, we introduce a novel model selection criterion, the 
least squares approximated information criterion (LIC). The LIC integrates two 
foundational concepts: the least squares approximation (LSA) of a loss function
and the quasi-likelihood under independence model criterion~(QIC). 
\citet{wang2007unified} proposed LSA as a simplified approximation 
to certain loss functions. We use LSA to approximate a loss function whose
gradient is the estimating functions for the joint mean, scale, and correlation
model. QIC, originally introduced by \citet{pan2001akaike}, has been 
extended by \citet{luo2022conditional} to accommodate multiple 
estimation equations.
Detailed descriptions of QIC and LSA can be found in 
Appendix~\ref{appendix:quasi} and Appendix~\ref{appendix:LSA}, respectively. 
Drawing inspiration from these two approaches, we have devised the LIC, which 
is a more general and robust model selection criterion.

The LIC is constructed by combining the quadratic term from LSA with the 
penalty term from QIC. Let $\hat{\theta}_n^f$ be the estimated parameter vector
of the full model and $\hat{\theta}_n^c$ be the estimated parameter vector of a
candidate model. Let $\tilde{\theta}_n^c$ be an augmented version of
$\hat\theta_n^c$ such that it has the same length as $\hat\theta_n^f$ with the
missing elements corresponding to unselected variables filled by zero.
The joint LIC is then defined as
\begin{equation}
  \label{eq:LICj}
  \text{LIC}_J (\tilde{\theta}_n^c)
  = (\tilde{\theta}_n^c -  \hat{\theta}_n^f)^{\top}
  \hat{\Sigma}_{1n}^f (\tilde{\theta}_n^c -  \hat{\theta}_n^f)
  + \log(n) \text{tr} (\hat{\Sigma}_{1n}^c \hat{V}_{n}^c),
\end{equation}
where $\hat{\Sigma}_{1n}^f$ and $\hat{\Sigma}_{1n}^c$ are
the estimated joint slope matrices of the joint estimating equations 
under the full and candidate model, respectively, and 
$\hat{V}_{n}^c$ is the estimated covariance matrix of 
$\hat\theta_n^c$. The inverse of the slope matrices $\Sigma_{1n}^f$ and
$\Sigma_{1n}^c$ are  the bread piece of the sandwich estimator under the full
and candidate model, and their detailed definitions are presented 
in Appendix~\ref{appendix:sand_var}.
Typically, it is recommended to keep the intercept in
the mean, scale, and correlation models.

The two terms in LIC have interpretations similar to those in existing criteria
with subtle differences. The first term measures the loss or lack-of-fit 
relative to the full model. It is the second-order approximation of
an objective function whose gradient is the estimating equations. The full 
model has lack-of-fit zero. The second term penalizes the complexity
of the candidate model. Note that the penalty term draws inspiration from the
correlation information criterion introduced by \citet{hin2009working}. It
accounts for the working correlation structures in the slope matrix. This is in
contrast to the corresponding term in QIC, which is the slope matrix under
working independence regardless of the working correlation structure being 
used. The penalty scale $\log(n)$, which corresponds to the Bayesian 
information criterion (BIC) penalty, could be replaced with 2 in the spirit of
Akaike information criterion (AIC) as formulated by \citet{pan2001akaike}.
Models with smaller LIC are preferred.
\citet{luo2022conditional} reported that the BIC-type penalty gave 
better model
selection results. As shown in our simulation studies, with the correct 
variance
estimator, the BIC-type penalty still selects the true models more frequently
than the AIC-type penalty.

The computational burden in an all-subset model selection for the joint model,
which requires fitting a total of $2^{p+r+q}$ candidate models, can be reduced
by a marginal version of the LIC. The marginal LIC selects each component of 
the
mean, scale, and correlation models separately, while the parameters of the 
other two components are fixed at their estimates from their corresponding full
models. That is, only one set of estimating
equations~\eqref{eq:beta}--\eqref{eq:gamma} is considered at a time. The slope
matrices $\Sigma_{1n}^f$ and $\Sigma_{1n}^c$ only contain the slope matrix of 
the component being considered in the full and candidate model, respectively. 
The variance matrix $\hat V_n^c$ becomes the variance matrix of the parameter 
estimator for the component being considered in the candidate model. The 
marginal
version of the LIC is similar to the approach used in the implementation of QIC
in \citet{luo2022conditional} (see Appendix~\ref{appendix:quasi}). 
With this strategy, we have only $2^{p}+2^{r}+2^{q}$ candidate models
under consideration.

The LIC offers distinct advantages over QIC in two crucial aspects. First, QIC 
is limited to performing marginal variable selections for mean, scale, and 
correlation models separately, whereas the LIC can be employed for joint model 
selection. Despite the computational complexity in all-subset model selection,
the possibility of joint model selection in all three components is an 
important
advantage. The LIC can be applied to other settings where one or multiple
estimating equations are used to estimate model parameters, and there is no 
natural
objective functions whose derivatives are the estimating equations. Further, 
QIC
relies on assuming working independence to construct the quasi-likelihood, 
while
LIC takes working correlation structure into account in the spirit of
\citet{hin2009working}.

\section{Simulation Studies}
\label{sec:sim_study}

Simulation studies were carried out with two primary objectives: to compare the
accuracies of the variance estimators in \citet{luo2022conditional} 
and \citet{yan2004estimating} and 
to demonstrate the effectiveness of the proposed LIC in model selection. The 
studies were designed under conditions similar to those outlined in the study 
by \citet{luo2022conditional}. Since their studies showed no sign of 
mismatch
between their empirical variance and estimated variance of all the parameter
estimators, we modified their correlation model so that
the mismatch is clearly observable, while the variance estimator of
\citet{yan2004estimating} performs well. For ease of referencing, 
we use LP and YF to denote the variance estimators from 
\citet{luo2022conditional} and \citet{yan2004estimating}, respectively.

\subsection{Simulation Design}

To highlight the differences in the approaches used by YF and LP, we introduced 
two significant alterations and one minor modification to the simulations as 
described in the study by \citet{luo2022conditional}. First, we 
introduced a 
non-constant variance function, thereby illustrating the potential advantages 
of modeling the scale parameter as opposed to directly modeling the variance. 
Second, we adopted a Toeplitz correlation structure, which enhances the 
significance of the lower triangular blocks within the sandwich estimator's 
bread matrix. Additionally, we made a slight change to simplify the matrices 
$R_{1i}$ and $R_{2i}$ to identity matrices, conforming to prior research that 
has indicated their negligible impact on the estimation process 
\citep{liang1986longitudinal, luo2022conditional}.

The explicit formulation of the model is provided as follows. Assuming the 
error vector, denoted as $e_{i} = (e_{i1}, \ldots, e_{im_{i}})$, is generated 
from a normal distribution $N_{m_i}(0, \Sigma_i)$, where $\Sigma_i$ represents 
the covariance matrix defined in equation~\eqref{eq:cov}, the adapted 
mean-scale-correlation model can be subsequently expressed as
\begin{align}
        \label{eq:simu_beta}
        Y_{ij} &= \beta_0 + x_{ij1} \beta_1 + x_{ij2} \beta_2 + e_{ij}, \\
        \label{eq:simu_lambda}
        \log \phi_{ij} &= \lambda_0 + z_{ij1} \lambda_1 + z_{ij2} \lambda_2, \\
        \label{eq:simu_gamma}
        \rho_{ijk} &= h_{ijk0} \gamma_0 + h_{ijk1} \gamma_1 + 
        h_{ijk2} \gamma_2,
\end{align}
where the variables $(x_{ij1}, x_{ij2})$ and $(z_{ij1}, z_{ij2})$ are 
identically distributed and drawn from $N_2(0, \text{CS}(0.5))$, a bivariate 
normal distribution 
with a mean of $0$ and a covariance matrix following a compound symmetry 
structure with a parameter value of $0.5$. By defining the covariate vectors as 
$h_{ijk} = (h_{ijk0}, h_{ijk1}, h_{ijk2})$ within the correlation model using 
indicator vectors, we establish a Toeplitz correlation structure to represent 
the within-cluster correlations. In this configuration, when $|j-k| = 1$, the 
covariate vector $h_{ijk}$ is equal to $(1, 0, 0)$, signifying that the 
correlation $\rho_{ijk}$ for two observations separated by a distance of 1 
within the same cluster is precisely $\gamma_0$. Similarly, when $|j-k| = 2$, 
the corresponding covariate vector $h_{ijk}$ becomes $(0, 1, 0)$, and 
$\rho_{ijk}$ is assigned the value $\gamma_1$. When $|j-k| = 3$, the associated 
covariate vector $h_{ijk}$ is $(0, 0, 1)$, and $\rho_{ijk}$ takes on the value 
$\gamma_2$.

The Toeplitz structure described above is a special case of the correlation 
model in equation~\eqref{eq:simu_gamma}. In this 
particular scenario, we have chosen a fixed cluster size of 4 due to the 
presence of only three covariates in the correlation model. However, in 
practice, the cluster size can vary, allowing for a more flexible correlation 
model. In subsequent simulation studies for parameter estimation and model 
selection, we will employ different correlation structures specified through 
different $h_{ijk}$ vectors, as well as varying parameters $\beta$, $\lambda$, 
and $\gamma$, and distinct variance functions $v_{ij}$ in the model in 
equations~\eqref{eq:simu_beta}--\eqref{eq:simu_gamma}.

\subsection{Parameter Estimation}

We began by comparing the accuracy and effectiveness of YF's and 
LP's methods for parameter estimation. This study includes two distinct 
simulation scenarios: one with a constant variance function, and the other 
incorporating a non-constant variance function. For both scenarios, identical 
parameter values are assigned: $\beta = (0, -1, 0.5)$, $\lambda = (2, 1, -1)$, 
and $\gamma = (0.5, 0.25, 0.125)$. It's noteworthy that with this correlation 
model, an AR(1) correlation matrix with a lag-1 correlation value set at $0.5$ 
is employed. Under the AR(1) correlation structure, 
the constraints detailed in \citet{luo2022conditional} to ensure the 
positive 
definiteness of generated correlation matrices $R_{1i}$ are not necessary. We 
generated 1000 replicates, each with $n = 300$ clusters.

We tabulated several results from this analysis. In addition to the point 
estimates (EST), we calculated the empirical standard errors (ESE) for the 300 
parameter estimates. We also calculated the average standard errors (ASE) 
across 300 sandwich variance estimates. To compare the 
performance of the two methods, we estimated the standard deviations using both 
the YF and LP approaches. The EST and ASE were then used to determine the 
coverage percentage (CP) for each parameter estimate. The coverage percentage 
represents the proportion of replicates in which the true value falls within 
the corresponding 95\% confidence interval. This measure allows us to evaluate 
the accuracy and reliability of the confidence intervals generated by the 
estimation procedure.

\begin{table}[tbp]
  \caption{
  Parameter estimation results for the mean-scale-correlation model on normally 
  distributed data with two variance functions, based on 1000 replicates, each 
  with 300 clusters.}
  \label{tab:sim_par_est}
  \centering
  \begin{tabular}{rrrrrrrrr}
    \toprule
    & \multicolumn{4}{c}{YF} & \multicolumn{4}{c}{LP}\\
    \cmidrule(lr){2-5}\cmidrule(lr){6-9}
    
    & \multicolumn{1}{c}{EST} & \multicolumn{1}{c}{ESE} & \multicolumn{1}{c}{ASE} 
    & \multicolumn{1}{c}{CP} & \multicolumn{1}{c}{EST} & \multicolumn{1}{c}{ESE} 
    & \multicolumn{1}{c}{ASE} & \multicolumn{1}{c}{CP} \\
    \midrule
    
    \addlinespace
    \multicolumn{9}{l}{\RNum{1}. $v_{ij} = 1$}\\
    
    $\beta_0$ & 0.006 & 0.116 & 0.115 & 94.8 & 0.115 & 0.006 & 0.116 & 94.8 \\
    $\beta_1$ & $-$0.996 & 0.064 & 0.065 & 94.9 & $-$0.996 & 0.064 & 0.065 & 94.9 
    \\
    $\beta_2$ & 0.497 & 0.065 & 0.065 & 94.7 & 0.497 & 0.065 & 0.065 & 94.7 \\
    $\lambda_0$ & 1.993 & 0.048 & 0.049 & 94.9 & 1.993 & 0.048 & 0.049 & 94.9 \\
    $\lambda_1$ & 1.002 & 0.048 & 0.046 & 93.2 & 1.002 & 0.048 & 0.046 & 93.3 \\
    $\lambda_2$ & $-$1.002 & 0.046 & 0.046 & 95.1 & $-$1.002 & 0.046 & 0.046 & 
    95.1  \\
   
    $\gamma_0$ & 0.500 & 0.028 & \textcolor{blue}{0.029} & \textcolor{blue}{96.0} 
    & 0.500 & 0.028 & \textcolor{blue}{0.047} & \textcolor{blue}{99.7} \\ 
    $\gamma_1$ & 0.249 & 0.041 & \textcolor{blue}{0.042} & \textcolor{blue}{95.4} 
    & 0.249 & 0.041 & \textcolor{blue}{0.048} & \textcolor{blue}{97.9} \\
    $\gamma_2$ & 0.127 & 0.057 & \textcolor{blue}{0.056} & \textcolor{blue}{94.1} 
    & 0.127 & 0.057 & \textcolor{blue}{0.058} & \textcolor{blue}{95.4} \\
    
    \addlinespace
    \multicolumn{9}{l}{\RNum{2}. $v_{ij} = 1 + 0.35 \times \tanh(\mu_{ij})$}\\
    
    $\beta_0$ & 0.007 & 0.113 & 0.115 & 94.8 & 0.007 & 0.113 & 0.115 &
    94.8\\
    $\beta_1$ & $-$0.996 & 0.065 & 0.064 & 94.3 & $-$0.996 & 0.064 & 0.064 &
    94.8\\
    $\beta_2$ & 0.497 & 0.069 & 0.067 & 94.6 & 0.496 & 0.069 & 0.067 &
    95.0\\
    $\lambda_0$ & \textcolor{blue}{1.991} & \textcolor{blue}{0.057} & 
    \textcolor{blue}{0.056} & \textcolor{blue}
    {94.4} & \textcolor{blue}{1.972} & \textcolor{blue}{0.049} & 
    \textcolor{blue}{0.049} & \textcolor{blue}{91.2}\\
    $\lambda_1$ & \textcolor{blue}{0.999} & \textcolor{blue}{0.049} &
    \textcolor{blue}{0.047} & \textcolor{blue}
    {93.7} & \textcolor{blue}{0.763} & \textcolor{blue}{0.049} & \textcolor{blue}
    {0.046} & \textcolor{blue}{0.10}\\
    $\lambda_2$ & \textcolor{blue}{$-$1.000} & \textcolor{blue}{0.048} &
    \textcolor{blue}{0.049} & \textcolor{blue}
    {95.3} & \textcolor{blue}{$-$0.882} & \textcolor{blue}{0.045} &
    \textcolor{blue}{0.047} & \textcolor{blue}{29.4}\\
    $\gamma_0$ & 0.499 & 0.028 & \textcolor{blue}{0.029} & \textcolor{blue}
    {96.0} & 0.499 & 0.028 & \textcolor{blue}{0.047} & \textcolor{blue}
    {99.5}\\
    $\gamma_1$ & 0.248 & 0.041 & \textcolor{blue}{0.042} & \textcolor{blue}
    {94.6} & 0.248 & 0.041 & \textcolor{blue}{0.048} & \textcolor{blue}
    {97.6}\\
    $\gamma_2$ & 0.127 & 0.058 & \textcolor{blue}{0.056} & \textcolor{blue}
    {94.5} & 0.126 & 0.058 & \textcolor{blue}{0.058} & \textcolor{blue}
    {95.0}\\
    \bottomrule
  \end{tabular}
\end{table}

Table~\ref{tab:sim_par_est} summarizes several key observations. In the first 
scenario, the responses $Y_{ij}$ follow a normal distribution with constant 
variance, and the scale parameter $\phi_{ij}$ is the same as the variance 
$\sigma^2_{ij}$. Both GEE approaches yield robust point estimates for mean, 
scale, and correlation parameters, ensuring accurate modeling of both the mean 
and covariance structures. Nevertheless, we observe that YF's method produces 
ASEs that are closer to the ESEs, particularly for the estimates highlighted in 
the table. In this case, the disparities between the two variance estimators 
are primarily evidenced by the correlation parameters, which is reasonable 
given that block $E_n$ of $\hat{\Sigma}_{1n}$ would have an expected value of 
0.354 rather than 0. Consequently, the presence of off-diagonal block $E_n$ in 
${\hat{\Sigma}_{1n}}$ leads to better variance estimators via YF, and 
subsequently provides confidence intervals with CP approaching 95\%.

Additional observations are presented in the second scenario. Apart from 
adopting a distinct correlation structure from \citet{luo2022conditional}, we 
introduced a variance function defined as $v_{ij} = 1 + 0.35 \times 
\tanh(\mu_{ij})$, resulting in function values uniformly distributed within the 
interval $(0.65, 1.35)$. Importantly, the model proposed by 
\citet{luo2022conditional} is not capable of handling data with 
non-constant 
variance function values $v_{ij}$ because they directly regressed on the 
variances $\sigma^2_{ij}$ rather than the scales $\phi_{ij}$, where 
$\sigma^2_{ij} = \phi_{ij}v_{ij}$. Given the limitations of LP's model, their 
point estimates of $(\lambda_0, \lambda_1, \lambda_2)$ deviate from true 
values. Thus, it should be noted that while comparing the performance of these 
two methods, it is important to consider that LP is not a correct model. 
Despite this inherent unfairness in comparison, we still note that the 
difference between the empirical standard errors and average standard errors of 
$(\lambda_0, \lambda_1, \lambda_2)$ highlights an error in the sandwich 
variance estimator of \citet{luo2022conditional} in this scenario. 
This error 
arises from the fact that blocks $B$ and $D$ do not have expected values of 0 
due to the variance function's dependence on $\mu_{ij}$. The bias in point 
estimates and variance estimates contributes to significantly low CP in the 
95\% confidence interval. In contrast, YF's method exhibits better performance 
in both variance and confidence interval estimations. Further details can be 
found in Appendix~\ref{appendix:coincidence}.

\subsection{Model Selection}

In this section, three different scenarios were used to
evaluate the effectiveness of the proposed model selection criterion LIC, in 
comparison with the existing QIC-based methods. In the first scenario, the 
model structure follows that employed by \citet{luo2022conditional} 
for the correlation model. Here, the value of $\rho_{itk}$ is parameterized as 
$f(\rho_{itk}) = h_{itk0} \gamma_0 + h_{itk1} \gamma_1 + h_{itk2} \gamma_2$, 
with $\gamma = (0.2, -0.2, 0)^{\top}$, and a rescaled Fisher’s z transformation 
link function. The vector
$h_{itk} = (h_{itk0}, h_{itk1}, h_{itk2})^{\top}$ is generated from a normal
distribution $N_3(0,\text{CS}(0.3))$, subject to the constraint that
\begin{equation*}
    \|h_{itk}\|_2 \leq \|\gamma\|_2^{-1} \min 
    \left\{ \left| f\left(-\frac{0.9}{m_i-1}\right)\right|, 
    \left| f\left(\frac{0.9}{m_i-1}\right) \right| \right\},
\end{equation*}
where $m_i \sim \text{Binomial}(10,0.7)$. 
The constraint is required for simulation studies to ensure the 
positive definiteness of $R_{1i}$. It is worth noting that when both $h_{itk0}$ 
and $h_{itk1}$ are generated with small scales, the magnitudes of 
$h_{itk0}\gamma_0$ and $h_{itk1}\gamma_1$ are comparable to the scale of 
$h_{itk0}\gamma_0 + h_{itk1}\gamma_1$. Consequently, distinguishing between the 
two candidate models $f(\rho_{itk}) = h_{itk0}\gamma_0$ and 
$f(\rho_{itk}) = h_{itk1}\gamma_1$ as well as identifying the true model for 
$\gamma$ can be challenging. Thus, the relatively poorer performance of 
correlation model selection in this scenario is considered acceptable, as 
discussed by \citet{luo2022conditional}. The second scenario is 
scenario 
\RNum{2} used in Section 4.1, where we consider different parameter vectors of 
$\beta=(1, -1, 0)^{\top}$, $\lambda= (2, 1, 0)^{\top}$, and 
$\gamma = (0.5, 0.5, 0)^{\top}$ with 300 clusters, each with a size of 4. The 
third model setup is similar to the second one but with a constant variance 
function.

In this numerical study, we evaluated the performance of four different model 
selection methods within the context of these three scenarios. The model 
selection methods under consideration include joint selection using LIC, 
marginal selection using LIC, model selection using QIC with the sandwich 
variance of YF, and model selection using QIC with the sandwich variance of LP. 
Furthermore, all four methods incorporated both the BIC-type penalty
and the AIC-type penalty  to assess 
the impact of larger and smaller penalties on model selection.

\begin{table}[tbp]
  \centering
  \caption{Percentage of correctly selected models based on 1000 replicates, each
    with 300 clusters.}
  \label{tab:simu_model_sel}
  \begin{tabular}{ccc ccc cccc}
    \toprule
    Scenario & \makecell{Penalty \\ Coefficient} & Method & $\beta$ & $\lambda$ & 
    $\gamma$ & $(\beta,\lambda)$ & $(\beta,\gamma)$ &
    $(\lambda$,$\gamma)$ & $(\beta,\lambda$,$\gamma)$
    \\
    \midrule
  
    \RNum{1} & $\log(n)$ & Joint & 95.9 & 96.9 & 83.2 & 92.8 & 79.2 & 80.2 & 76.2 
    \\
      &     & Marginal & 97.8 & 98.1 & 86.3 & 95.8 & 84.1 & 84.7 & 82.5 \\
      &     & QIC (YF) & 97.5 & 98.4 & 86.3 & 95.9 & 84.1 & 84.9 & 82.7 \\ 
      &     & QIC (LP) & 97.5 & 98.4 & 86.0 & 95.9 & 83.8 & 84.6 & 82.4 \\  
      & 2 & Joint & 80.3 & 82.8 & 82.9 & 66.9 & 66.1 & 67.7 & 54.3 \\
      &     & Marginal & 81.7 & 83.5 & 84.0 & 68.7 & 68.5 & 69.3 & 56.9 \\
      &     & QIC (YF) & 82.2 & 84.0 & 84.0 & 69.1 & 68.8 & 69.8 & 57.4 \\ 
      &     & QIC (LP) & 82.2 & 84.0 & 84.1 & 69.1 & 68.9 & 69.9 & 57.5 \\ 
      
    \addlinespace
    \RNum{2} & $\log(n)$ & Joint & 98.0 & 96.5 & 95.3 & 94.9 & 93.5 & 92.2 & 90.6 
    \\ 
      &     & Marginal & 98.3 & 96.6 & 98.0 & 95.2 & 96.3 & 94.6 & 93.2 \\ 
      &     & QIC (YF) & 94.8 & 97.0 & 97.9 & 92.1 & 92.7 & 94.9 & 90.1 \\ 
      &     & QIC (LP) & 94.8 & 96.6 & 97.9 & 91.7 & 92.7 & 94.4 & 89.6 \\ 
      & 2 & Joint & 83.9 & 81.3 & 80.2 & 68.3 & 67.5 & 66.0 & 55.6 \\ 
      &     & Marginal & 84.0 & 81.9 & 81.3 & 69.0 & 68.5 & 67.1 & 56.8 \\ 
      &     & QIC (YF) & 79.3 & 82.4 & 81.2 & 65.4 & 65.0 & 67.1 & 54.0 \\ 
      &     & QIC (LP) & 79.3 & 81.9 & 82.3 & 65.1 & 65.5 & 67.7 & 54.2 \\ 
      
    \addlinespace
    \RNum{3} & $\log(n)$ & Joint & 97.3 & 97.9 & 95.3 & 95.3 & 92.9 & 93.4 & 91.0 
    \\ 
      &     & Marginal & 97.6 & 97.9 & 97.9 & 95.5 & 95.5 & 95.8 & 93.5 \\ 
      &     & QIC (YF) & 94.6 & 97.9 & 97.6 & 92.5 & 92.2 & 95.5 & 90.1 \\
      &     & QIC (LP) & 94.6 & 97.9 & 97.9 & 92.5 & 92.5 & 95.8 & 90.4 \\
      & 2 & Joint & 82.6 & 82.1 & 81.4 & 68.3 & 67.2 & 66.4 & 55.3 \\ 
      &     & Marginal & 83.2 & 82.6 & 83.8 & 69.2 & 69.2 & 68.6 & 57.2 \\
      &     & QIC (YF) & 76.8 & 82.9 & 83.7 & 63.6 & 63.8 & 68.7 & 52.2 \\
      &     & QIC (LP) & 76.8 & 83.0 & 84.0 & 63.7 & 64.0 & 69.2 & 52.5 \\
  
    \bottomrule
  \end{tabular}
\end{table}

Table~\ref{tab:simu_model_sel} provides a comprehensive summary of the results 
obtained from various model selection criteria. The primary overarching 
observation is that the BIC-type penalty consistently outperforms the AIC 
penalty across all criteria and scenarios. This consistent pattern suggests 
that the BIC penalty is more effective in identifying significant variables, 
while the AIC penalty tends to favor more complex models. In Scenario \RNum{1}, 
our proposed marginal selection using LIC yields results comparable to those 
obtained with the QIC methods. Although the joint selection using LIC exhibits 
a slightly lower correct model selection probability compared to the other 
three methods, it 
offers a more versatile approach with acceptable performance. In the case of 
the Toeplitz correlation model with a non-constant variance function (scenario 
\RNum{2}), both our proposed joint and marginal selection using LIC perform 
well in distinguishing true models, achieving correct selection rates of 
over 95\% for mean, scale, and correlation models, with overall correct 
selection rates exceeding 90\%. 
Specifically, LIC demonstrates superior performance in 
the mean model, which is reasonable given our model's Toeplitz correlation 
structure, making the assumption of working independence in QIC inappropriate. 
The third scenario involving the Toeplitz correlation model with a constant 
variance function exhibits a similar pattern to scenario \RNum{2}, with 
slightly improved performance, particularly for $\lambda$. This improvement is 
expected, as the scale model has a lower variance and higher coverage rate. The 
proposed LIC for model selection achieves good coverage rates of 
96.7\%, 95.5\%, and 97.3\% for scenarios \RNum{1}, \RNum{2}, and \RNum{3}, 
respectively. 
Additionally, we observed that QIC (YF) and QIC (LP) produce nearly identical 
results. This similarity is anticipated because the variance function used in 
this simulation study, $v_{ij} = 1 + 0.35 \times \tanh(\mu_{ij})$, restricts 
the range to $(0.65, 1.35)$. Given that this range is close to 1, its impact 
on the QIC is minimal, leading to the similar performance observed between the 
two QIC methods. Therefore, it is expected that the two QIC methods would yield 
comparable results.


Overall, the results indicate that the proposed LIC criterion is effective in 
both joint and marginal model selection across a wide range of model setups, 
including cases 
with complex correlation structures and constant/non-constant variance 
functions. The performance of the proposed LIC is promising in terms of 
correctly identifying the true models for the mean, scale, and correlation 
coefficients of the joint model.

\section{Uterine Artery Pulsatility Index Data}
\label{sec:PI_data}

We now apply the proposed GEE method to the Uterine Artery (UtA) Pulsatility 
Index (PI) data, in which each individual can be treated as a cluster with 
multiple measurements. Both parameter estimation and model selection strategies 
are employed in this dataset. 
Preeclampsia (PE) is the most common cause of illness and mortality in pregnant 
women, fetuses, and newborns worldwide. In a prospective cohort study conducted
by \citet{song2019first}, Doppler velocimetry was used to measure PI, 
which has 
been found to be useful in predicting early-onset PE (pre-eclampsia before 34 
weeks of pregnancy). A total of 255 patients were included in the study, with 
each patient having two different measurement methods (manual/machine) for PI 
values recorded on both sides of the uterus (left/right). For each method-side 
combination, there were three repeated measurements, resulting in a total of 12 
measurements for each pregnant woman. As suggested by \citet{song2019first}, 2 
patients with gestational hypertension, 3 with late-onset PE, and 3 cases of 
placental centralization were removed. There were 247 patients left, for a 
total number of 2964 observations in the dataset.

Our objective is to model the mean-scale-correlation structures of the PI 
values, which range from 0.34 to 4.88 with a mean of 1.44 and a standard 
deviation of 0.58. Previous research by \citet{song2019first} found 
that the 
left-side UtA had a higher PI than the right-side UtA on the side without the 
placenta. To accomplish a similar goal, we created two categorical variables: 
method (0 for manual and 1 for machine) and side (0 for right and 1 for left). 
Other variables, such as age, crown-rump length (CRL), gestational age in weeks 
(week), and placental laterality (PL, 0 for right and 1 for left), were kept 
the same across the 12 measurements within each cluster. Additionally, the 
interaction between side and PL was included as a variable, indicating whether 
the placenta and the measurement were on the same side or not, and how it could 
affect the PI values. We use this design matrix for the mean and scale models, 
with identity and log links, respectively.

\begin{figure}[tbp]
    \centering
    \includegraphics[width = 0.8\textwidth]{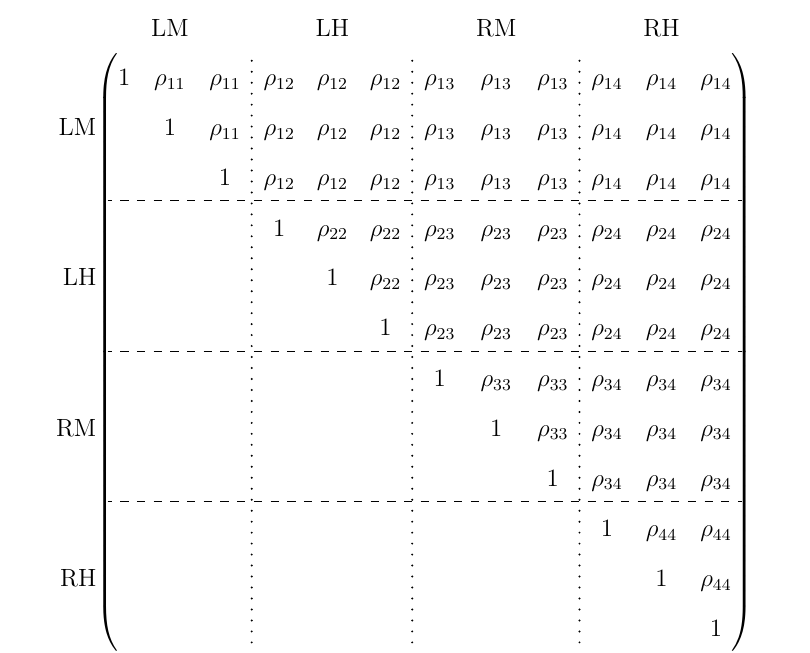}
    \caption{The correlation matrix constructed in the analysis of UtA PI 
    data.}
    \label{fig:real_data_cor_matrix}
\end{figure}

For the correlation model, we constructed a design matrix that includes 10 
indicator variables to account for the different types of correlation. There 
are 4 method-side combinations, and each combination can be correlated with 
itself and 3 other combinations, resulting in a total of 10 types of 
correlation $(\rho_{11}, \rho_{12}, \ldots, \rho_{44})$, which were presented 
as the $12 \times 12$ correlation matrix in 
Figure~\ref{fig:real_data_cor_matrix}. Each block in the matrix denoted as LM, 
LH, RM, and RH corresponds to three measurements of the left-machine, 
left-human, right-machine, and right-human, respectively. In each cluster, 
there are a total of 66 correlations, resulting in a grand total of 16,302 
correlations. The correlation model incorporates 10 types of correlations 
within each cluster, which correspond to 10 covariates included in the model. 
We defined LM-LM as the correlation between two left-machine measurements and 
LM-RM to represent the correlation between a left-machine measurement and a 
right-human measurement, and so on for the other types of correlations. We set 
LM-LM and LM-RM as the reference levels, indicating whether the correlation is 
between two measurements on the same side or different side. This allows us to 
represent the variations of the remaining eight types of correlations with 
respect to their reference levels. For example, LM-LH represents the difference 
between the LM-LH correlation and its reference level, SS, while LM-RH 
correlation represents the difference between the LM-RH correlation and its 
reference level, DS. In this setting, the correlation link is specified as 
identity.

\begin{table}[tbp]
  \centering
  \caption{Generalized estimating equation result (parameter estimates, standard
  errors, and model selection) for the model fitted for 247 patients and 2964
  observations in the UtA PI data.}
  \label{tab:real_full_model}
  \begin{tabular}{crrrrrrrr}
      \toprule
      & \multicolumn{2}{c}{\multirow{2}{*}{Full Model}} & 
      \multicolumn{6}{c}{Selected Model} \\
      \cmidrule(lr){4-9} & & & \multicolumn{2}{c}{QIC (YF)} & 
      \multicolumn{2}{c}{QIC (LP)} & \multicolumn{2}{c}{LIC} \\
      \cmidrule(lr){2-3}\cmidrule(lr){4-5}\cmidrule(lr){6-7}\cmidrule(lr){8-9}
      Variables & \multicolumn{1}{c}{EST} & \multicolumn{1}{c}{SE} & 
      \multicolumn{1}{c}{EST} & \multicolumn{1}{c}{SE} & \multicolumn{1}{c}{EST}
      & \multicolumn{1}{c}{SE} & \multicolumn{1}{c}{EST} & \multicolumn{1}{c}{SE} 
      \\
      \midrule
      \multicolumn{9}{c}{Mean model:}\\
      Intercept & 1.181 & 0.039 & 1.179 & 0.039 & 1.179 & 0.039 & 1.179 & 0.039\\
      Age & 0.006 & 0.007 &  &  \\
      CRL & $-$0.012 & 0.006 & $-$0.011 & 0.005 & $-$0.011 & 0.005 & $-$0.011 & 
      0.005 \\
      Week & $-$0.003 & 0.067 &  &  \\
      PL & 0.273 & 0.061 & 0.282 & 0.061 & 0.281 & 0.061 & 0.282 & 0.061 \\
      Method & 0.161 & 0.007 & 0.163 & 0.008 & 0.163 & 0.008 & 0.163 & 0.008 \\
      Side & 0.411 & 0.039 & 0.409 & 0.040 & 0.410 & 0.040 & 0.410 & 0.040 \\
      PL $\times$ Side & $-$0.647 & 0.059 & $-$0.651 & 0.059 & $-$0.650 & 0.059 & 
      $-$0.651 & 0.059\\
      [1em]
      \multicolumn{9}{c}{Scale model:}\\
      Intercept & $-$1.439 & 0.106 & $-$1.387 & 0.076 & $-$1.294 & 0.066 & 
      $-$1.294 & 0.066 \\
      Age & 0.007 & 0.016 &  &  \\
      CRL & $-$0.021 & 0.013  \\
      Week & $-$0.041 & 0.169 &  &  \\
      PL & 0.103 & 0.149 & &  \\
      Method  & 0.230 & 0.038 & 0.239 & 0.039 & 0.243 & 0.041 & 0.243 & 0.041 \\
      Side & 0.280 & 0.128 & 0.246 & 0.112 &  &  &  &  \\
      PL $\times$ Side & $-$0.539 & 0.172 & $-$0.453 & 0.140 & $-$0.302 & 0.112 & 
      $-$0.302 & 0.112\\
      [1em]
      \multicolumn{9}{c}{Correlation model:}\\
      Same Side & 0.830 & 0.023 & 0.850 & 0.013 & 0.850 & 0.056 & 0.850 & 0.013  
      \\
      LM-LH & 0.020 & 0.016 &  &  & & \\
      LH-LH & 0.013 & 0.028 &  &  \\
      RM-RM & $-$0.031 & 0.034 &  &  \\
      RM-RH & 0.031 & 0.027 &  &  &  &  \\
      RH-RH & 0.058 & 0.031 &  &  &  & \\
      Different Side & 0.471 & 0.046 & 0.484 & 0.047 & 0.482 & 0.059 & 0.482 & 
      0.046  \\
      LM-RH & 0.023 & 0.015 &  &  \\
      LH-RM & $-$0.002 & 0.016 &  &  \\
      LH-RH & 0.032 & 0.021 &  &  & &  \\
      \bottomrule
  \end{tabular}
\end{table}

Table~\ref{tab:real_full_model} first displays the outcomes of our model 
fitting in the context of parameter estimations (EST) and their corresponding 
estimated standard errors (SE) within the full model, which align with the 
findings of previous research by \citet{song2019first}. These 
findings confirm that PI values for 
measurements taken on the left side are higher compared to the right side. 
Additionally, the location of measurement, particularly on the placental side, 
exerts a clear influence on PI values, with measurements on the placental side 
exhibiting decreased values due to a negative interaction coefficient. 
Moreover, the influence of PL on PI values is statistically significant, with 
the left placenta associated with elevated PI values. Some additional findings 
other than those by \citet{song2019first} are revealed in the results. 
Specifically, an inverse relationship between CRL and PI values is shown, 
indicating that an increase in CRL is linked to a reduction in PI. 
Comparatively, machine-measured PI values tend to be higher when contrasted 
with manual measurements. In terms of the scale model, machine-acquired 
measurements and those taken on the left side exhibit larger scales, resulting 
in increased variance under a constant variance function assumption. 
Conversely, measurements on the placental side demonstrate smaller scales, 
supported by a negative coefficient, indicating reduced variance compared to 
non-placental measurements. In the correlation model, only the two reference 
levels within each cluster exhibit significance, with the correlation between 
two measurements on the same side at approximately 0.83 and that on different 
sides at around 0.47. It is worth noting that the methodology employed for 
measurements does not substantially influence observed correlations, which 
forms the basis for subsequent model selection procedures.

The model selection results obtained by three different methods are presented 
in Table~\ref{tab:real_full_model}. First, we included the outcomes of two 
approaches employing QIC as the selection criteria. The first approach, denoted 
as QIC (YF), utilized the variance estimator introduced by 
\citet{yan2004estimating}, while the second approach, labeled QIC 
(LP), employed the variance estimators proposed by \citet{luo2022conditional}. 
The standard errors of the model selected by QIC (LP) were computed using LP's 
variance estimator, whereas the standard errors for the other two selected 
models were calculated using YF's variance estimator. For LIC, we explored both 
joint selection and marginal selection. It's worth noting that we did not 
perform a joint model selection simultaneously for all the mean, scale, and 
correlation models 
since this would require fitting $2^{23}$ models due to the total of 26 
variables in the model, which is computationally infeasible. Consequently, we 
selected the mean model and scale model jointly, while keeping the correlation 
model fixed at the full model, with separate model selection for the 
correlation model. Interestingly, both joint and marginal LIC 
criteria resulted in the selection of the same model. Consequently, we have 
presented the results only once in Table~\ref{tab:real_full_model}.

Table~\ref{tab:real_full_model} shows that all three methods select the same 
mean model. All three methods select intercept, CRL, PL, method, side, and 
interaction as the significant variables. The selected covariates exhibit 
relatively minor changes from the full model. Notably, the conclusion that Age 
and Week are not significant is substantiated by the outcomes presented in the 
full model, as evidenced by the relatively large estimated standard errors 
associated with these two variables.

The model selection results differ in the scale model. While all methods 
identify the intercept, method, and interaction as significant variables in the 
scale model, QIC~(LP) and LIC do not include the side. This is because 
block $B_n$ in the sandwich estimator, as detailed in 
Appendix~\ref{appendix:coincidence}, is not precisely equal to 0, even though 
its expected value is 0 as presumed in the LP estimator.
In the scale model selected by QIC (YF), notable discrepancies can be observed, 
particularly in the estimated intercept and interaction, which displays a 
significant difference from the full model alongside a substantially reduced 
SE. The other variables exhibit EST and SE close to the full model. Similar 
results can be found in the scale model selected by LIC and QIC (LP), with an 
even more apparent difference in the EST of the intercept and interaction, and 
a more substantial reduction in their SE. Notably, QIC (YF) successfully 
identifies the additional variable that both LIC and QIC (LP) fail to select.

In the scale model chosen using QIC (YF), noticeable discrepancies are evident, 
particularly in the ESTs of intercept and interaction, which exhibit 
significant differences from the full model, along with substantially reduced 
standard errors (SE). Other variables in this model show ESTs and SEs that are 
close to those in the full model. Similar findings are observed in the scale 
model selected by LIC and QIC (LP), with even more apparent differences in the 
ESTs of intercept and interaction, and a more substantial reduction in their 
SEs. It is noteworthy that QIC (YF) successfully identifies an additional 
variable that both LIC and QIC (LP) do not select.

The results of the correlation model selection are particularly intriguing, as 
all methods concur in selecting the same side and different side as the only 
two significant variables. In essence, this finding strongly suggests that the 
correlation between two PI measurements for one patient depends solely on 
whether they are measured on the same side or not. Specifically, the 
correlation between two measurements on the same side is markedly higher than 
that for measurements on different sides. In this context, the 
measurement method does not appear to play a significant role in the 
correlation model. In the selected model, the standard errors obtained using 
LP's formula are considerably larger than those obtained using YF's formula. 
This observation serves as additional evidence that YF's variance estimator is 
more versatile and robust.

\section{Discussion}
\label{sec:discussion}

The joint mean-variance-correlation model of \citet{luo2022conditional} is a
special case of that of \citet{yan2004estimating}. Both works used 
the same set
of estimating equations but presented different variance estimators of the
regression coefficient estimator. Our analysis unveiled scenarios where the
variance estimator of \citet{luo2022conditional} may be misleading. 
This issue 
arises due to the hierarchical structure within the estimating equations, 
leading to the inappropriate use of a block-diagonal matrix in their sandwich 
variance estimator's bread component. When off-diagonal blocks substantially 
deviate from zero, this can affect the accuracy of the variance estimator. The 
same oversight was made by \citet{kastner1999comparison} and
\citet{ziegler2000familial} in a two-equation 
setup. The variance estimator of \citet{yan2004estimating} is shown 
to be 
dependable and efficient, as implemented in the R package \texttt{geepack}
\citep{hojsgaard2006r}, as demonstrated through both simulation studies and 
real-data applications.

The nature of the data can significantly impact our methods' performance. 
First, the GEE approach may not be valid when data are not missing completely 
at random (MCAR), meaning that the probability of missingness depends on the 
responses \citep{laird1988missing,liang1986longitudinal}. Missing data can also 
reduce efficiency, and complicate model convergence. Addressing these 
challenges requires methods like weighting \citep{robins1995analysis} and 
imputation \citep{paik1997generalized} for valid and reliable results. Second, 
for
longitudinal data, methods can be simplified due to the specific correlation 
structure for each subject. Our simulation study and real data analysis 
used the Toeplitz correlation structure, simplifying the correlation 
model. The commonly used AR(1) or CS correlation structures have only one
parameter. Working correlation structures could be constructed to allow, for
example, block structures among multivariate longitudinal data.
Third, selecting the correct correlation 
structure and link function is vital for accurate GEE modeling with discrete 
data like binary or Poisson responses \citep{fitzmaurice1993regression}. A 
proper correlation structure ensures unbiased parameter estimates, and the 
link function must align with the data type, such as using a logit link for 
binary outcomes. Additionally, diagnostics and model checking are crucial 
to validate the model's assumptions and performance. Finally, measurement
errors in covariates bring extra challenges in GEE modeling with the need to
correct the biases in regression coefficient estimation
\citep{yi2012functional, lau2022bias}. The consequences on the scale and
correlation parameter estimation merit further studies.

Model selection with LIC appears to be a promising approach 
for general modeling setups where estimating equations are used which may not 
correspond to the derivatives of any objective function. In our
simulation studies under the joint mean-variance-correlation modeling, it
exhibits generally good performance in comparison with the QIC approach. The
capability of joint selection for all parameters across all estimating 
equations makes it a unique contribution to model selection based on estimating
equations. Between the joint and marginal versions, there are possibilities to
consider certain subsets of components jointly. A limitation is that the LIC is 
not designed to handle high-dimensional covariates, as addressed by 
\citet{wang2007unified}. It extends the AIC, BIC, and QIC in standard
applications where the number of covariates is small to moderate. Further work 
is needed to extend this approach to the high-dimensional setting.

\section*{Acknowledgements}
The authors are grateful for the email discussions and MATLAB  code from
Drs. Jianxin Pan and Renwen Luo.

\begin{appendices}
\section{Sandwich Variance Estimator}
\label{appendix:sand_var}

According to \citet{yan2004estimating}, the vector 
$[n^{1/2}(\hat\beta_n - \beta)^{\top},
n^{1/2}(\hat\lambda_n - \lambda)^{\top}, 
n^{1/2}(\hat\gamma_n - \gamma)^{\top}]^{\top}$ 
has an asymptotic normal distribution characterized by a mean 
of zero and a covariance matrix with a sandwich structure
$\Sigma_1^{-1} \Sigma_2 \{\Sigma_1^{-1}\}^{\top}$, where
$\Sigma_1$ and $\Sigma_2$ are the limit that two random
matrices converge to in probability, respectively, as defined next.

Matrix~$\Sigma_1$ is the limit of $n^{-1}\Sigma_{1n}$, where
$\Sigma_{1n}$ is the slope
matrix of the estimation equations~\eqref{eq:beta}--\eqref{eq:gamma}, 
evaluated at $\hat\theta_n$. In particular,
\begin{equation*}
  \Sigma_{1n} = 
  \begin{pmatrix}
    A_n & 0 & 0 \\
    -B_n & C_n & 0 \\
    -D_n & -E_n & F_n
  \end{pmatrix}
\end{equation*}
with
\begin{equation*}
    \begin{aligned}
        A_n &= \sum_{i=1}^n D_{1i}^{\top} V_{1i}^{-1}
        D_{1i},&&
        B_n = \sum_{i=1}^n D_{2i}^{\top} V_{2i}^{-1}
        \partial s_{i} / \partial\beta^{\top},&&
        C_n = \sum_{i=1}^n D_{2i}^{\top} V_{2i}^{-1}
        D_{2i}, \\
        D_n &= \sum_{i=1}^n D_{3i}^{\top} V_{3i}^{-1}
        \partial z_{i} / \partial\beta^{\top},&&
        E_n = \sum_{i=1}^n D_{3i}^{\top} V_{3i}^{-1}
        \partial z_{i} / \partial\lambda^{\top},&&
        F_n = \sum_{i=1}^n D_{3i}^{\top} V_{3i}^{-1}
        D_{3i},
    \end{aligned}
\end{equation*}
and the undefined partial derivatives can be derived as
\begin{equation*}
    \begin{aligned}
        \partial s_{ij} / \partial \beta =& \frac{1}{v_{ij}^2}\left[ -2D_{1ij}
        \epsilon_{ij} v_{ij} - \epsilon_{ij}^2 \frac{\dd v_{ij}}{\dd
        \mu_{it}}D_{1ij} \right],\\
        \partial z_{ijk} / \partial \beta =& \frac{1}{\sqrt{\phi_{ij}v_{ij}
        \phi_{ik}v_{ik}}}  \left[ -D_{1ij} \epsilon_{ik} -
        D_{1ik}\epsilon_{ij} - \frac{1}{2} \epsilon_{ij} \epsilon_{ik} \left(
        \frac{\dd v_{ij}}{\dd \mu_{ij}} D_{1ij} \frac{1}{v_{ij}} + 
        \frac{\dd v_{ik}}
        {\dd \mu_{ik}} D_{1ik} \frac{1}{v_{ik}} \right) \right],\\
        \partial z_{ijk} / \partial \lambda =& 
        - \frac{\epsilon_{ij} \epsilon_{ik}}
        {2 \sqrt{\phi_{ij} v_{ij} \phi_{ik} v_{ik}}} \left( D_{2ij} \frac{1}
        {\phi_{ij}} + D_{2ik} \frac{1}{\phi_{ik}} \right),
    \end{aligned}
\end{equation*}
where $\epsilon_{ij} = Y_{ij} - \mu_{ij}$ and 
$\epsilon_{ik} = Y_{ik} - \mu_{ik}$. Note that the partial derivatives
$\partial z_{ijk} / \partial \beta$ and $\partial z_{ijk} / \partial \lambda$
here correct a previous misprint in \citet{yan2004estimating}.
It is reassuring 
that the R package \texttt{geepack} \citep{hojsgaard2006r} has consistently
implemented these calculations correctly.

Matrix~$\Sigma_2$ is the limit of $n^{-1}\Sigma_{2n}$
where $\Sigma_{2n}$ is the covariance matrix of the three stacked estimating
equations~\eqref{eq:beta}--\eqref{eq:gamma}. This covariance matrix
can be constructed as 
\begin{equation*}
  \sum_{i=1}^n
  \begin{pmatrix}
    D_{1i}^{\top} V_{1i}^{-1} \hat{\cov}(Y_i) V_{1i}^{-1} D_{1i} &
    D_{1i}^{\top} V_{1i}^{-1} \hat{\cov}(Y_i, s_i) V_{2i}^{-1} D_{2i} &
    D_{1i}^{\top} V_{1i}^{-1} \hat{\cov}(Y_i, z_i) V_{3i}^{-1} D_{3i} \\
    D_{2i}^{\top} V_{2i}^{-1} \hat{\cov}(s_i, Y_i) V_{1i}^{-1} D_{1i} &
    D_{2i}^{\top} V_{2i}^{-1} \hat{\cov}(s_i) V_{2i}^{-1} D_{2i} &
    D_{2i}^{\top} V_{2i}^{-1} \hat{\cov}(s_i, z_i) V_{3i}^{-1} D_{3i} \\
    D_{3i}^{\top} V_{3i}^{-1} \hat{\cov}(z_i, Y_i) V_{1i}^{-1} D_{1i} &
    D_{3i}^{\top} V_{3i}^{-1} \hat{\cov}(z_i, s_i) V_{2i}^{-1} D_{2i} &
    D_{3i}^{\top} V_{3i}^{-1} \hat{\cov}(z_i) V_{3i}^{-1} D_{3i}
  \end{pmatrix}.
\end{equation*}

Then, the joint covariance matrix of the estimated GEE parameters 
$\hat\theta_n$, can be estimated by a sandwich variance estimator
$\hat{V}_n = \Sigma_{1n}^{-1} \Sigma_{2n} \{\Sigma_{1n}^{-1} \}^{\top}$.

\section{Coincidence in Sandwich Estimators}
\label{appendix:coincidence}

The sandwich variance estimator proposed by \citet{luo2022conditional} is
different from that of \citet{yan2004estimating}. However, both 
methods
would result in almost the same result by following the simulation study in
\citet{luo2022conditional}. To find out the reason behind this, the 
difference between these two methods is studied in detail here. From
Appendix~\ref{appendix:sand_var}, the sandwich variance estimator is
$\hat{\Sigma}_{1n}^{-1} \hat{\Sigma}_{2n} \{\hat{\Sigma}_{1n}^{-1}\}^{\top}$,
and the first matrix $\hat{\Sigma}_{1n}$ is a block lower triangular matrix.
On the other hand, the sandwich formula used in \citet{luo2022conditional}
estimates the variance of $\hat{\beta}_n$, $\hat{\lambda}_n$, and
$\hat{\gamma}_n$
independently, which is equivalent to set $\{\hat{\Sigma}_{1n}\}$ as a block
diagonal matrix. It is straightforward that blocks $B_n$, $D_n$, and $E_n$ 
should be further investigated.

Now consider the simulation setting of \citet{luo2022conditional}.
The response variables are normally distributed and the
variance is not a function of the mean. In terms of the framework in
\citet{yan2004estimating}, for the second equation, the scale is 
the equivalent
to the variance in that of \citet{luo2022conditional}, and the 
variance function
is just the constant 1, i.e., $\dd v_{ij} / \dd \mu_{ij} = 0$. In this case,
\begin{align*}
    \partial s_{ij} / \partial \beta
    =& -2\epsilon_{ij} D_{1ij},\\
    \partial z_{ijk} / \partial \beta
    =& \frac{1}{\sqrt{\phi_{ij} \phi_{ik}}}\left(
    -D_{1ij} \epsilon_{ik} - D_{1ik} \epsilon_{ij} \right),\\
    \partial z_{ijk} / \partial \lambda =&
    -\frac{\epsilon_{ij} \epsilon_{ik}}{2 \sqrt{\phi_{ij} \phi_{ik}}}
    \left( D_{2ij} \frac{1}{\phi_{ij}} + D_{2ik} \frac{1}{\phi_{ik}} \right).
\end{align*}
While the first two terms have expectations zero, the third one does not:
\begin{align}
    \label{eq:E3}
    \E[\partial z_{ijk} / \partial \lambda ]
    =&-\frac{1}{2}\E\left[\frac{\epsilon_{ij} \epsilon_{ik}}
    {\sqrt{\phi_{ij} \phi_{ik}}}\right]
    \left( D_{2ij} \frac{1}{\phi_{ij}} + D_{2ik} \frac{1}{\phi_{ik}} \right).
\end{align}
The consequence is that the off-diagonal blocks of $\Sigma_{1n}$ do not
necessarily disappear as $n\to\infty$. We have $B_n / n $ and
$D_n / n$ converging to zero, but not $E_n / n$ in general.

\begin{figure}[tbp]
    \centering
    \includegraphics[width = \textwidth]{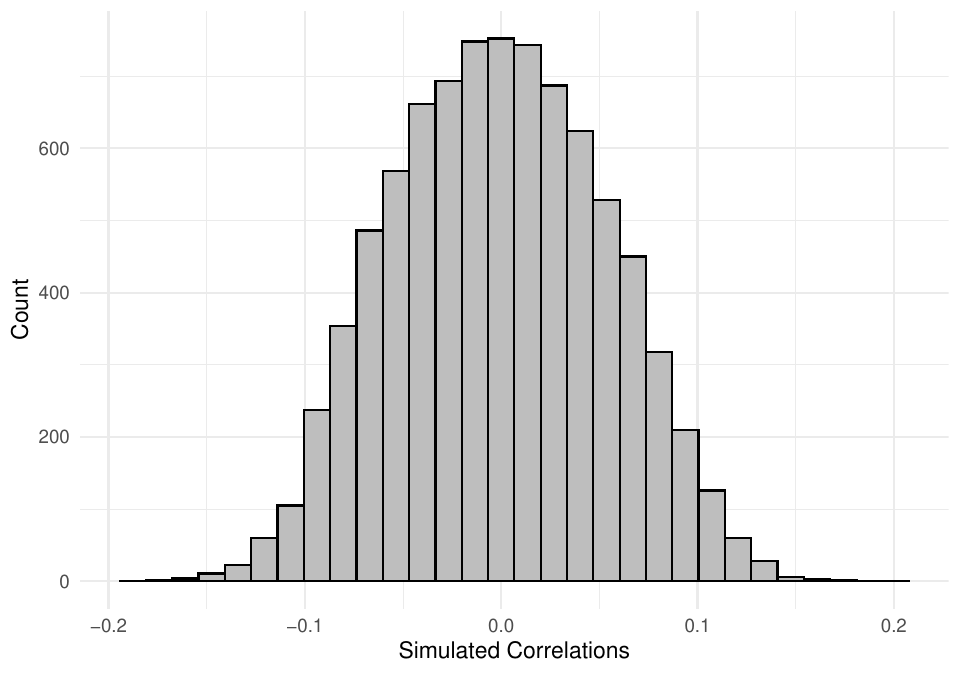}
    \caption{Histogram of the simulated correlations $\rho_{ijk}$ in LP's 
    simulation 
    study, where $X_{3ijk}$ was generated from $N_3(0, \text{CS}(0.3))$ and
    $\gamma$ was set to $(0.1, -0.2, 0.15)^\top$.}
    \label{fig:hist-delta}
\end{figure}

There are certain specific scenarios where $E_n / n \to 0$, in which case, the
variance estimator from \citet{luo2022conditional} would give 
correct result. The expectation in Equation~\eqref{eq:E3} is
$\rho_{ijk} = \cor(Y_{ij}, Y_{ik})$. When this expectation averaged over
all $i$, $j$, and $k$ is close to zero, $E_n$ would be close to zero.
In the simulation setting of \citet{luo2022conditional}, the true 
values of $\rho_{ijk}$'s can be calculated by
$g_3^{-1}(X_{3ijk} \gamma)$ from Equation~\eqref{eq:model}. 
Figure~\ref{fig:hist-delta} shows the histogram of these true correlation
parameters where $X_{3ijk}$ was generated from $N_3(0, \text{CS}(0.3))$ and
$\gamma$ was set to $(0.1, -0.2, 0.15)^\top$. The
distribution of $\rho_{ijk}$'s exhibits a 
central and symmetrical pattern around~0.
This explains why the flawed variance estimator of \citet{luo2022conditional} 
had good performance in their simulation studies.
In the general case where $E_n / n$ does not approach zero, as illustrated in
our simulation study, their variance estimate does not agree with the
empirical variance.

\section{QIC and Quasi-likelihoods}
\label{appendix:quasi}

In order to select for variance and correlation models, 
\citet{luo2022conditional} extended the QIC criterion proposed by 
\citet{pan2001akaike} into three 
model selection procedures. They also integrated the penality of the BIC,
which applies a stronger penalty to free parameters and 
favors a more parsimonious model. Let $\hat{\beta}_n^f$, $\hat{\lambda}_n^f$, 
and $\hat{\gamma}_n^f$ be the full model estimates, and $\hat{\beta}_n^c$, 
$\hat{\lambda}_n^c$, and $\hat{\gamma}_n^c$ be the candidate model estimates. 
The quasi-likelihoods for the mean, scale, and correlation models are defined 
as 
\begin{equation*}
    \begin{aligned}
        Q_1(\hat{\beta}_n^c, \hat\lambda_n^f, \hat\gamma_n^f) &= 
        \sum_{i=1}^n \sum_{j=1}^{m_i} \int_{y_{ij}}^{\hat\mu_{ij}} 
        \frac{y_{ij} - t}{\tilde\phi_{ij} v(t)} \, \dd t, \quad 
        &&\hat{\mu}_{ij} = g_1^{-1} (X_{1ij}^\top \hat{\beta}_n^c), \\
        Q_2(\hat{\beta}_n^f, \hat\lambda_n^c, \hat\gamma_n^f) &= 
        \sum_{i=1}^n \sum_{j=1}^{m_i} \int_{\tilde s_{ij}}^{\hat\phi_{ij}} 
        \frac{\tilde s_{ij} - t}{2t^2} \, \dd t, \quad 
        &&\hat{\phi}_{ij} = g_2^{-1} (X_{2ij}^\top \hat{\lambda}_n^c),\\
        Q_3(\hat{\beta}_n^f, \hat\lambda_n^f, \hat\gamma_n^c) &= 
        \sum_{i=1}^n \sum_{j=1}^{m_i-1} \sum_{k=j+1}^{m_i} 
        \int_{\tilde z_{ijk}}^{\hat\rho_{ijk}} 
        \frac{\tilde z_{ijk} - t}{1+t^2} \, \dd t, \quad 
        &&\hat{\rho}_{ijk} = g_3^{-1} (X_{3ijk}^\top \hat{\gamma}_n^c),\\
    \end{aligned}
\end{equation*}
with $\tilde\phi_{ij}$, $\tilde s_{ij}$, and $\tilde z_{ijk}$ being evaluated 
at $\hat{\beta}_n^f$, $\hat{\lambda}_n^f$, and $\hat{\gamma}_n^f$. 
Subsequently, their model selection criteria are formulated as
\begin{align*}
    \text{QIC}(\hat\beta_n^c) &= 
    -2Q_1(\hat{\beta}_n^c, \hat\lambda_n^f, \hat\gamma_n^f) + 
    \log(n) \text{tr}(\hat{\Omega}_{\beta n}^c \hat{V}_{\beta n}^c), 
    \quad \hat{\Omega}_{\beta n}^c = 
    \frac{-\partial^2 Q_1(\beta, \hat\lambda_n^f, \hat\gamma_n^f)}  
    {\partial \beta \beta^{\top} }
    \Biggr\rvert_{\beta = \hat{\beta}_n^c},\\
    \text{QIC}(\hat\lambda_n^c) &= 
    -2Q_2(\hat{\beta}_n^f,\hat\lambda_n^c, \hat\gamma_n^f) + 
    \log(n) \text{tr}(\hat{\Omega}_{\lambda n}^c \hat{V}_{\lambda n}^c), 
    \quad \hat{\Omega}_{\lambda n}^c = 
    \frac{-\partial^2 Q_2(\hat{\beta}_n^f, \lambda, \hat\gamma_n^f)}  
    {\partial \lambda \lambda^{\top} }
    \Biggr\rvert_{\lambda = \hat{\lambda}_n^c},\\
    \text{QIC}(\hat\gamma_n^c) &= 
    -2Q_3(\hat{\beta}_n^f, \hat\lambda_n^f, \hat\gamma_n^c) + 
    \log(n) \text{tr}(\hat{\Omega}_{\gamma n}^c \hat{V}_{\gamma n}^c), 
    \quad \hat{\Omega}_{\gamma n}^c = 
    \frac{-\partial^2 Q_3(\hat{\beta}_n^f, \hat\lambda_n^f, \gamma)} 
    {\partial \gamma \gamma^{\top} }
    \Biggr\rvert_{\gamma = \hat{\gamma}_n^c}, 
\end{align*}
and $\hat{V}_{\beta n}^c$,
$\hat{V}_{\lambda n}^c$, and $\hat{V}_{\lambda n}^c$ can be estimated by the 
corresponding parts of the sandwich variance estimator.

The model selection process was carried out 
separately for mean, variance, and correlation. This means that during the 
selection of the mean model, $\lambda$ and $\gamma$ were retained as the full 
model estimates $\hat{\lambda}_n^f$ and $\hat{\gamma}_n^f$ respectively. The 
candidate model with the lowest QIC was then chosen as the best mean model. 
Analogous procedures were followed for the selection of variance and scale 
models.

\section{Least Squares Approximation}
\label{appendix:LSA}

\citet{wang2007unified} proposed a method that uses LSA as a 
simple approximation to the original loss function for achieving a unified and 
straightforward estimation of the least absolute shrinkage and selection 
operator (LASSO). Assume that there is a loss function $L_n(\theta)$ with a 
continuous second-order derivative with respect to~$\theta$. This assumption is 
shown to be independent of the implementation of
LSA and is only used to illustrate the idea. 
By a second-order Taylor
expansion at $\hat\theta_n$,
\begin{equation}
\label{eq:LSA}
    n^{-1}L(\theta)\approx n^{-1}L_n(\hat\theta_n) + 
    n^{-1} \dot{L}_n(\hat\theta_n)^{\top} (\theta - \hat\theta_n) + 
    \frac{1}{2}(\theta - \hat\theta_n)^{\top}\left\{ 
    \frac{1}{n} \ddot{L}_n(\hat\theta_n) \right 
    \}(\theta - \hat\theta_n), 
\end{equation}
where $\dot{L}_n(\hat\theta_n)$ and $\ddot{L}_n(\hat\theta_n)$ are the first- 
and second-order derivatives of the loss function. Since $\hat\theta_n$ is a 
minimizer of $L_n(\hat\theta_n)$, we know that $\dot{L}_n(\hat\theta_n)=0$. 
Moreover, we can ignore $n^{-1}L_n(\hat\theta)$ since it is a constant. 
Therefore, $n^{-1}L(\theta)$ can be approximated by a quadratic term
\begin{equation*}
    \frac{1}{2}(\theta - \hat\theta_n)^{\top}\left\{ \frac{1}{n}
    \ddot{L}_n(\hat\theta_n) \right\}(\theta - \hat\theta_n).
\end{equation*}

In the framework of GEE1, if we assume the quasi-likelihood function 
$Q(\hat{\beta}_n^c)$ is well-defined and $\hat\beta_n^f \approx \beta$, we 
can approximate $2Q(\hat{\beta}_n^c)$ by expanding equation~\eqref{eq:LSA} 
at $\hat\beta_n^f$ as
\begin{equation*}
    (\hat\beta_n^c - \hat\beta_n^f)^{\top}\left\{
    \ddot{L}_n(\hat\beta_n^f) \right\}(\hat\beta_n^c - \hat\beta_n^f).
\end{equation*}
where $\ddot{L}_n(\hat\beta_n^f)$ can be computed as the bread matrix 
$n\hat\Sigma_{1n}$ in the sandwich variance estimator because 
$n\hat\Sigma_{1n}$ is the slope matrix of the estimating equations, and the 
estimating equations can be viewed as the first-order slope matrix of an 
unknown loss function. This is different from \citet{wang2007unified}, where 
the inverse of the estimated covariance matrix, $\hat V_{\hat\beta_n^f}^{-1}$, 
is used to estimate $\ddot{L}_n(\hat\beta_n^f)$. In this case, assuming 
$Y_{ij}$ follows a normal distribution, the quadratic term reduces to the log-
likelihood. Consequently, our LIC defined in equation~\eqref{eq:LICj} reduces 
to the BIC, and it reduces to the AIC if the coefficient of the penalty term is 
set to~$2$.

\end{appendices}

\bibliography{references}
\bibliographystyle{chicago}

\end{document}